\documentclass[12pt,twoside]{article}
\usepackage{epsfig,float,wrapfig}
\usepackage{color}
\usepackage{fancyhdr}

\usepackage{amsmath}
\usepackage{amssymb}

\newcommand{\be}{\begin{equation}}
\newcommand{\ba}{\begin{eqnarray}}
\newcommand{\ea}{\end{eqnarray}}

\def\a{\alpha}

\def\d{\delta}
\def\e{\epsilon}

\def\j{\psi}

\def\m{\mu}
\def\n{\nu}

\def\p{\pi}
\def\q{\theta}
\def\r{\rho}
\def\s{\sigma}

\def\F{\Phi}
\def\G{\Gamma}

\def\S{\Sigma}

\def\cs{{\cal S}}

\newcommand{\ti}{\tilde}

\newcommand{\pa}{\partial}
\newcommand{\tr}{\mbox{tr}}
\newcommand{\qb}{\bar{\q}}
\newcommand{\fb}{\bar{\F}}

\newcommand{\var}[2]{\frac{\delta \Phi(#1)}{\delta \Phi(#2)}}

\renewcommand{\title}[1]{\null\vspace{25mm}\noindent{\Large{\bf #1}}\vspace{10mm}}
\newcommand{\authors}[1]{\noindent{\large #1}\vspace{20mm}}
\newcommand{\address}[1]{{\center{\noindent #1\vspace{0mm}}}}
\renewcommand{\abstract}[1]{\vspace{17mm}
\noindent{\small{\em Abstract.} #1}\vspace{2mm}}

\DeclareMathAlphabet\mathbb  {U}{msb}{m}{n}
\DeclareFontFamily{U}{msb}{} \DeclareFontShape{U}{msb}{m}{n}{
  <5> <6> <7> <8> <9> gen * msbm
  <10> <10.95> <12> <14.4> <17.28> <20.74> <24.88> msbm10
  }{}

\makeatletter
\def\section{\@startsection{section}{1}{\z@}{-3.25ex plus -1ex minus
             -.2ex}{1.5ex plus .2ex}{\normalfont\bfseries}}
\def\subsection{\@startsection{subsection}{1}{\z@}{-3.25ex plus -1ex
                minus -.2ex}{1.5ex plus .2ex}{\normalfont\itshape}}

\renewenvironment{thebibliography}[1]
         {\section*{References}\frenchspacing\small
          \begin{list}{[\arabic{enumi}]}
         {\usecounter{enumi}\parsep=2pt\topsep 0pt
         \settowidth{\labelwidth}{[#1]}
         \leftmargin=\labelwidth\advance\leftmargin\labelsep
         \rightmargin=0pt\itemsep=0pt\sloppy}}{\end{list}}

\makeatother
\begin{document}   \setcounter{table}{0}
 
\begin{titlepage}
\begin{center}
\hspace*{\fill}{{\normalsize \begin{tabular}{l}
                              {\sf hep-th/0007050}\\
			      {\sf  TUW 00-22}\\
			      {\sf  UWThPh-2000-29.}
                              \end{tabular}   }}

\title{\vspace{5mm} The Superfield Formalism Applied to the Noncommutative Wess-Zumino Model}

\vspace{20mm}

\authors {  \Large{ A. A. Bichl$^{1}$, J. M. Grimstrup$^{2}$, H. Grosse$^{3}$, L. Popp$^{4}$,\\ M. Schweda$^{5}$, R. Wulkenhaar$^{6}$ }}    \vspace{-20mm}

\vspace{10mm}
       
\address{$^{1,2,4,5}$  Institut f\"ur Theoretische Physik, Technische Universit\"at Wien\\
      Wiedner Hauptstra\ss e 8-10, A-1040 Wien, Austria}
\address{$^{3,6}$  Institut f\"ur Theoretische Physik, Universit\"at Wien\\Boltzmanngasse 5, A-1090 Wien, Austria   }
\footnotetext[2]{Work supported by The Danish Research Agency.}
\footnotetext[4]{Work supported in part by ``Fonds zur F\"orderung der Wissenschaftlichen Forschung'' (FWF) under contract P13125-PHY and P13126-PHY.}
\footnotetext[6]{Marie-Curie Fellow.}       
\end{center} 
\thispagestyle{empty}
\begin{center}
\begin{minipage}{12cm}

\vspace{10mm}

{\it Abstract.} We introduce the notion of superoperators on noncommutative $\mathbb{R}^4$ and re-investigate in the framework of superfields the noncommutative 
Wess-Zumino model as a quantum field theory. In a highly efficient manner we are able to confirm the result that this model is renormalizable to all orders.

\end{minipage}\end{center}
\end{titlepage}

\section{Introduction}

There are no doubts that the concept of space-time as a differentiable
manifold cannot be extrapolated to extremely short distances
\cite{snyder}. Simple heuristic arguments combining the principles of
both general relativity and quantum theory imply that it is impossible
to locate a particle with an arbitrarily small uncertainty \cite{dfr}.
This means that standard differential geometry is certainly not an
adapted method for physics at short distances. On the other hand, our
standard description of fundamental interactions is exclusively based
upon standard differential geometry -- fibre bundles for the standard
model and Riemannian geometry for gravity.

If standard differential geometry is not appropriate -- what else
should replace it? A promising candidate is \emph{noncommutative
  geometry} pioneered by Alain Connes \cite{ac}, see \cite{c} for a
recent review. An excellent book \cite{gbvf} on this subject will soon
appear. Noncommutative geometry is the attempt to extend the
principles of quantum mechanics to geometry itself: The use of
operator algebras, Hilbert spaces, functional analysis. Within this
framework the analogue of gauge theory has been developed which
reduces to Yang-Mills theory if the geometry is commutative. The
first example of such a theory on a noncommutative space appeared
almost ten years ago, when Connes and Rieffel developed classical two
dimensional Yang-Mills theory on the noncommutative torus \cite{cr}.
There is also an interesting class of ``almost commutative'' geometries
which allow to treat Yang-Mills and Higgs fields on an equal footing
and lead to a new understanding of spontaneous symmetry breaking, see
e.g.\ \cite{cisk}.  The analogue of external field quantization on
noncommutative spaces was proposed in \cite{vgm}.

Thus, the strategy of noncommutative geometry is to generalize the
mathematical structures encountered in experimentally confirmed
physics. Another approach to short-distance physics is string theory/
M theory which tries to guess physics from new first principles. At
first sight it seems unlikely that noncommutative geometry and string
theory could be related. However, it has been shown that certain
noncommutative geometries arise as limiting cases of string
theory. The first hint came from
\cite{cds} where compactifications of M theory on the noncommutative
torus were introduced, leading to the interpretation that the step
from the commutative to the noncommutative torus corresponds to
turning on a constant background 3-form $C$. Then, in \cite{dh}
it was shown that this situation is obtained by starting with a type
IIa superstring theory with non-zero Neveu-Schwarz $B$ field and
taking a scaling limit according to \cite{s1,s2}. That idea was
thoroughly investigated in \cite{sw}. Using the results of \cite{ns}
about instantons on noncommutative $\mathbb{R}^4$, Seiberg and Witten
argued that there is an equivalence between the Yang-Mills theories on
standard $\mathbb{R}^4$ and $\mathbb{R}^4_{nc}$.

It should be mentioned that matrix theories were studied long before M
theory was proposed, and that these matrix theories did contain
certain noncommutative features. These models live on a lattice, and
the number of degrees of freedom is reduced when the size $N$ of the
matrix goes to infinity \cite{ek}. Putting them on a torus instead of a
lattice, twisted boundary conditions \cite{tHooft} are possible. Then
the action can be rewritten in terms of noncommuting 
matrix derivatives $[\Gamma^{(j)},\;.\;]$, with
$[\Gamma^{(2j)},\Gamma^{(2j+1)}]=-2\pi\mathrm{i}/N$, see \cite{altes}.

The Seiberg-Witten paper \cite{sw} inspired numerous attempts to
formulate quantum field theories on noncommutative geometries. Nevertheless,
quantum field theory on noncommutative spaces is also interesting in
its own right. As standard quantum field theory is the art to deal
with problems of interactions at short distances, see e.g.\ the
proceedings \cite{vw}, one should expect interesting features when
doing quantum field theory on spaces with different short-distance
structure. Singularities in standard quantum field theories are a
consequence of the point-like interactions. There has been some hope that
smearing out the points it is possible to avoid these ultraviolet
divergences. Example of geometries where points are replaced by some
sort of cells are the fuzzy spaces, see e.g.\ \cite{madore,gkp}. Such fuzzy
spaces also arise as limits of brane dynamics \cite{ars}.

That divergences are not avoided on $\mathbb{R}^4_{nc}$ was first noticed by
Filk \cite{f}. He showed that the noncommutative model contains
Feynman graphs which are identical with their commutative
counterparts. The $\mathbb{R}^4_{nc}$ is defined by the following
commutator of the coordinate operators $\{q^\mu\}$:
\[
[q^\mu,q^\nu] = \mathrm{i} \S^{\mu\nu}~,\qquad
\S^{\mu\nu} = -\S^{\nu\mu} \in \mathbb{R}~.
\]
Integrals corresponding to Feynman graphs in noncommutative QFTs
differ from their commutative counterparts by phase factors
$\mathrm{e}^{\frac{\mathrm{i}}{2}\S^{\mu\nu} p_\mu k_\nu}$, where $p,k$ are 
internal or external momenta. The case $p=k$ is possible, and in this
situation the integrals of the commutative and the noncommutative
theory coincide. 

This raised the question whether the noncommutative QFT is
renormalizable. On the one-loop level this was affirmed for
Yang-Mills theory on $\mathbb{R}^4_{nc}$ \cite{ms} and the
noncommutative 4-torus \cite{kw} as well as for supersymmetric
Yang-Mills theory in $(2+1)$ dimensions, with space being the 
noncommutative 2-torus \cite{sj}. Quantum electrodynamics on
$\mathbb{R}^4_{nc}$ was treated in \cite{h} and the BF-Yang-Mills
theory in \cite{b}.

These results lead to the hope that Yang-Mills theory on 
$\mathbb{R}^4_{nc}$ is renormalizable to all orders in perturbation theory.
It was shown however by Minwalla, Van Raamsdonk and Seiberg \cite{mvs}
that at least for scalar theories ($\phi^4$ on 
$\mathbb{R}^4_{nc}$ and $\phi^3$ on $\mathbb{R}^6_{nc}$) there is
a new type of infrared divergences which ruins the perturbative
renormalization beyond one loop. This follows immediately from the
work of Filk \cite{f}, it was nevertheless completely unexpected: The
oscillatory factors $\mathrm{e}^{\frac{\mathrm{i}}{2}\S^{\mu\nu} p_\mu
  k_\nu}$ render in four dimensions an otherwise (superficially)
divergent integral convergent if both $p,k$ are internal momenta. If
e.g.\ $p$ is external and $k$ is internal, the integral is convergent
as well, but of course only as long as $p \neq 0$, where the
ultraviolet divergence of the commutative theory reappears. This
manifests as an infrared divergence coming from a
ultraviolet-dangerous integration (UV/IR mixing). It turns out that
the power counting degree of the new superficial IR divergence of the
noncommutative theory coincides with the degree of the superficial UV
divergence of the commutative theory \cite{chr}.

Inspite of some rumour in the literature that Yang-Mills theory has
only logarithmic divergences, it turned out that Yang-Mills on
$\mathbb{R}^4_{nc}$ has quadratic IR divergences for the gluon
propagator and linear IR divergences for the 3-gluon vertex \cite{mst}
which prevent the perturbative renormalization. This result was also
derived from the scaling limit of string theory \cite{bcr,gkmrs}.
Graphs made exclusively of nested ghost propagator corrections have
only logarithmic divergences and are renormalizable at any loop order
\cite{gkw}. In \cite{mst} it was also shown that adding fermions in
the adjoint representation cancels the quadratic and linear IR
divergences at one loop. This was a hint that supersymmetric Yang
Mills could be renormalizable, which is also strongly supported by the
divergence analysis in \cite{chr}.

The UV/IR mixing was also observed in noncommutative complex scalar
$\phi^4$ theory \cite{abk} where the interaction potential
$a\phi^*\phi\phi^*\phi+b\phi^*\phi^*\phi\phi$ is only one-loop
renormalizable for $b=0$ or $a=b$. UV/IR mixing does also occur in
non-relativistic models \cite{gll}. Anomalies were investigated in
\cite{gbm} and the operator product expansion in \cite{z}. Discrete
symmetries (CPT) were investigated in \cite{sj2}.

A completely finite model is Chern-Simons theory on $\mathbb{R}^3_{nc}$
\cite{bgps}. Examples of models where the limit $\S^{\mu\nu} \to
0$ is smooth can be found in \cite{bksy}. Yang-Mills with fermions on
$\mathbb{R}^3_{nc}$ was studied in \cite{chu} with respect to the
induced Chern-Simons action. Noncommutative topological massive Yang-Mills was treated in \cite{Chen}. 
There are also interesting
two dimensional noncommutative models such as the nonlinear $\sigma$ model
\cite{dkl} and the noncommutative Wess-Zumino-Witten model \cite{msc}. 

Even at the tree-level a field theory on $\mathbb{R}^4_{nc}$ shows
unusual features such as violation of causality \cite{sst} and
S-matrix unitarity \cite{gm} if time is noncommutative. Experimental
limits on $\S^{\mu\nu}$ are discussed in \cite{mpr}. 
Beside the renormalizability problem, a quadratic IR divergence in the
one-loop gluon propagator leads to a confinement of the model to a
region which size is of the order $\sqrt{|\S^{\mu\nu}|}$. Thus,
such a model cannot have the standard model as its low energy
limit. The standard model can only be extended to noncommutative
space-time if the quadratic IR divergences cancel, which is probably
the case of supersymmetric versions. This motivates the interest in 
supersymmetry on noncommutative spaces, apart from the scaling limit of
superstring theory.

Concerning supersymmetry, in \cite{fl} a deformation also of the
anticommutator of the fermionic superspace coordinates $\theta$ was
considered, it was shown however that such a deformation is not
compatible with supertranslations and chiral superfields. This result
was also derived in \cite{cz} via the scaling limit of string theory.
A superspace formulation (at the classical level) of the Wess-Zumino
model and super Yang-Mills was given in \cite{t}. Employing the
component formulation it was eventually proved in \cite{ggrs} that the
Wess-Zumino (WZ) model on $\mathbb{R}^4_{nc}$ is renormalizable to all
orders in perturbation theory. One-loop renormalizability of $N=2$
super Yang-Mills was obtained in \cite{abkr}.

In this paper we extend the work of Filk \cite{f} to the superfield formalism and 
apply the techniques to the noncommutative Wess-Zumino model.
The paper is organized as follows: In section 2, following closer 
the work of Filk than \cite{fl,cz,t}, we introduce the notion of
a superoperator on noncommutative space-time. In section 3 the Wess-Zumino model
in the superfield formalism is introduced at the classical level, and
in section 4 we postulate a noncommutative version of the Gell-Mann Low formula, and
apply it to compute the one loop self-energy contributions. To be complete, we 
deduce the superfield Feynman rules in section 5, and apply them to the one loop vertex
corrections. In section 6 we discuss the renormalizability to all loop orders.

\arraycolsep 0.1em

\section{Superfields with noncommutative coordinates}
Following \cite{f,dfr} we consider the space-time coordinates of a flat space as self-adjoint operators in a Hilbert space with the following algebra
\be
\left[ q^{\m},q^{\n}\right]={\rm{i}}\S^{\m\n},\,\,\,\,\,\,\left[\S^{\m\n},q^{\r}  \right]=0,
\end{equation}
where $\S^{\m\n}$ is real and antisymmetric. In order to describe the superfields consistently within the framework of Filk one defines the operator
\be
T(k)=e^{{\rm{i}}k_{\m}q^{\m}},
\end{equation}
with the properties
\ba
T^{+}(k)&=&T(-k),\label{1a}\\
T(k)T(k')&=&T(k+k')e^{-{\rm{i}}k\times k'},\,\,\,\,\,\,\,\,\,k\times k'=\frac{1}{2}\S^{\m\n}k_{\m}k'_{\n},
\ea
and formally
\ba
\tr \left[T(k)\right]&=&\int d^{4}q\, e^{{\rm{i}}k_{\m}q^{\m}}\, =\, \prod_{\m}\d (k_{\m})(2\p)^4.\label{2a}
\ea
Additionally, we are dealing with classical chiral and anti-chiral superfields $\F(x,\q,\qb)$ and $\bar{\F}(x,\q,\qb)$ in the real representation \cite{piq}, defined on an ordinary manifold by 
\ba
\F(x,\q,\qb)&=&\F_{1}(x-{\rm{i}}\q\s\qb,\q)=e^{-{\rm{i}}\q\s\qb\pa}\F_{1}(x,\q),\\
\bar{\F}(x,\q,\bar{\q})&=&\fb_{2}(x+{\rm{i}}\q\s\qb,\qb)=e^{{\rm{i}}\q\s\qb\pa}\bar{\F}_{2}(x,\qb),
\ea
where
\ba
\F_{1}(x,\q)&=&A(x)+\q^{\a}\j_{\a}(x) + \q^{\a}\q_{\a}F(x),\\
\bar{\F}_{2}(x,\qb)&=&\bar{A}(x)+\qb_{\dot{\a}}\bar{\j}^{\dot{\a}}(x) + \qb_{\dot{\a}}\qb^{\dot{\a}}\bar{F}(x).
\ea
The corresponding covariant derivatives and further technical material on the superfield formalism is collected in appendix A.
To a chiral classical superfield $\F_{1}(x_{1},\q_{1})\equiv\F_{1}(1)$ one defines the Fourier-transform as\footnote{We use the notation $dx=d^{4}x$, $dk=\frac{d^{4}k}{(2\p)^{4}}$}
\be
\F_{1}(1)=\int dp\, e^{{\rm{i}}px_{1}}\ti{\F}_{1}(p,\q_{1}),
\end{equation}
where the coefficients of $\ti{\F}_{1}(p,\q_{1})$ belong to $\cs\left({\mathbb{R}^{4}}\right)$. The ``inverse'' reads
\ba
\ti{\F}_{1}(p,\q_{1})&=&\int dS_{2}\,e^{-{\rm{i}}px_{2}}\d(\q_{12})\F_{1}(x_{2},\q_{2})\nonumber\\
&=&\int dx_{2}\,e^{-{\rm{i}}px_{2}}\F_{1}(x_{2},\q_{1}),\label{4a}
\ea
where $dS_{2}$ and $\d(\q_{12})$ are given in appendix A. Corresponding to Filk \cite{f}, one associates to the classical chiral superfield $\F_{1}(x_{1},\q_{1})$ the following superoperator
\ba
\F_{1}(q_{1},\q_{1})&=&\int dS_{2}\int dk\, T(k)e^{{-\rm{i}}kx_{2}}\d(\q_{12})\F_{1}(x_{2},\q_{2})\nonumber\\
&=&\int dx_{2}\int dk\, T(k)e^{-{\rm{i}}kx_{2}}\F_{1}(x_{2},\q_{1})\nonumber\\
&=&\int dk\, T(k)\ti{\F}_{1}(k,\q_{1}).
\ea
The trace operation allows to recover the classical chiral superfield $\F_{1}(x_{1},\q_{1})$
\be
\F_{1}(1)=\int dk\, e^{{\rm{i}}kx_{1}}\tr\left[ \F_{1}(q_{1},\q_{1})T^{+}(k) \right].
\end{equation}
Following Filk's idea we are now able to construct a $\star$-product, the so-called Moyal product, of two classical superfields
\be
(\F_{1}\star \F_{1})(1)=\int dk\, e^{{\rm{i}}kx_{1}}\tr\left[\F_{1}(q_{1},\q_{1})\F_{1}(q_{1},\q_{1})T^{+}(k)  \right].\label{3a}
\end{equation}
With the relations (\ref{1a})-(\ref{2a}) eq.(\ref{3a}) becomes
\be
(\F_{1}\star \F_{1})(1)= \int dk_{1}\,\int dk_{2} e^{{\rm{i}}\left(k_{1}+k_{2}\right)x_{1}}e^{-{\rm{i}}k_{1}\times k_{2}}\ti{\F}_{1}(k_{1},\q_{1})\ti{\F}_{1}(k_{2},\q_{1}),\label{1}
\end{equation}
and with (\ref{4a}) there follows
\ba
(\F_{1}\star \F_{1})(1)&=&\int dk_{1}\,\int dk_{2}e^{{\rm{i}}(k_{1}+k_{2})x_{1}}e^{-{\rm{i}}k_{1}\times k_{2}}\nonumber\\&&
\times\int dS_{1'}\int dS_{2'}\,e^{-{\rm{i}}k_{1}x_{1'}-{\rm{i}}k_{2}x_{2'}}\d(\q_{11'})\d(\q_{12'})\F_{1}(x_{1'},\q_{1'})\F_{1}(x_{2'},\q_{2'}).
\ea
Additionally, one has also
\be
(\F_{1}\star \F_{1}\star \F_{1})(1)=\int dk\, e^{{\rm{i}}kx_{1}}\tr\left[ \F_{1}(q_{1},\q_{1})\F_{1}(q_{1},\q_{1})\F_{1}(q_{1},\q_{1})T^{+}(k) \right].\label{2}
\end{equation}
Repeating the same steps as before, eq.(\ref{2}) may be rewritten as
\ba
(\F_{1}\star \F_{1}\star \F_{1})(1)&=&\int dk_{1}\int dk_{2}\int dk_{3}\,e^{{\rm{i}}(k_{1}+k_{2}+k_{3})x_{1}}e^{-{\rm{i}}\sum_{i<j}^{3}k_{i}\times k_{j}}\tilde{\F}_{1}(k_{1},\q_{1})\tilde{\F}_{1}(k_{2},\q_{1})\tilde{\F}_{1}(k_{3},\q_{1})\nonumber\\
&=&\int dk_{1}\int dk_{2}\int dk_{3}\,e^{{\rm{i}}(k_{1}+k_{2}+k_{3})x_{1}}e^{-{\rm{i}}\sum_{i<j}^{3}k_{i}\times k_{j}}\nonumber\\&&\times
\int dS_{1'} \int dS_{2'} \int dS_{3'}\,e^{-{\rm{i}}k_{1}x_{1'}-{\rm{i}}k_{2}x_{2'}-{\rm{i}}k_{3}x_{3'}}\nonumber\\&&\times
\d(\q_{11'})\d(\q_{12'})\d(\q_{13'})
\F_{1}(x_{1'},\q_{1'})\F_{1}(x_{2'},\q_{2'})\F_{1}(x_{3'},\q_{3'}).
\ea
The $\F_{1}^{\star 3}$ is required to describe the corresponding interactions of the WZ-model.
With the functional derivative for chiral superfields
\be
\frac{\d \F_{1}(1)}{\d \F_{1}(2)}=\d_{S}(1,2)=\d(\q_{12})\d(x_{1}-x_{2})= -\frac{1}{4}\q_{12}^{2}\d(x_{1}-x_{2}),
\end{equation}
one checks that the above definitions imply also
\be
\frac{\d}{\d \F_{1}(2)}\int dS_{1}\,\left(\F_{1}\star\F_{1}\star\F_{1} \right)(1) = 3(\F_{1}\star\F_{1})(2).
\end{equation}
Finally, from (\ref{1}) we get the useful relation
\be
\int dS_{1}\,\left(\F_{1}\star \F_{1}\right)(1) =\int dS_{1}\,\F_{1}(1)\F_{1}(1) .
\end{equation}

\section{Superfield formulation of the noncommutative Wess-Zumino model at the tree level}
In four dimensional Minkowskian space-time the Wess-Zumino model in terms of superfields is defined at the tree level by the following action \cite{piq}
\ba
\G^{(0)}&=& \G^{(0)}_{kin}+\G^{(0)}_{m}+\G^{(0)}_{I} \nonumber\\
&=& \frac{1}{16}\int dV\, \fb\F + \frac{m}{8}\left[\int dS\, \F^{2}+ \int d\bar{S}\,\fb^{2}\right] 
+ \frac{g}{48}\left[\int dS \,\F^{\star 3}+ \int d\bar{S}\,\fb^{\star 3}\right].
\ea
Since $\F$ and $\fb$ are chiral and anti-chiral superfields the kinetic part of $\G^{(0)}$ can be rewritten as
\be
\G^{(0)}_{kin}=\frac{1}{16}\int dS\, \bar{D}^{2}\fb(x,\q,\qb)\F(x,\q,\qb) = \frac{1}{16}\int d\bar{S}\,\fb(x,\q,\qb) D^{2}\F(x,\q,\qb).\label{1b}
\end{equation}
Carrying out the $\q$ and $\qb$ integration, which is in fact a differentiation, one gets always the last component of a superfield (products of superfields are again superfields) which furnishes the component formulation of supersymmetric field models.
In order to derive the corresponding superfield propagators one uses the Legendre transformation between the functional for connected Green functions $Z_{c}$ and the vertex functional $\G^{(0)}$. Introducing external chiral and anti-chiral sources $J$ and $\bar{J}$ we have
\be
Z_{c}[J,\bar{J}]=\G^{(0)}[\F,\fb] + \int dS J\F + \int d\bar{S} \bar{J}\fb,
\end{equation}
with
\be
\frac{\d\G^{(0)}}{\d\F}=-J, \,\,\,\,\,\,\,\frac{\d\G^{(0)}}{\d\fb}=-\bar{J},\label{2b}
\end{equation}
and
\be
\frac{\d Z_{c}}{\d J}=\F,\,\,\,\,\,\,\,\frac{\d Z_{c}}{\d \bar{J}}=\fb.\label{3b}
\end{equation}
Solving eq.(\ref{2b}) for $\F=\F[J,\bar{J}]$ and $\fb=\fb[J,\bar{J}]$ one gets the desired superfield propagators
\ba
\left\langle T \F(1)\F(2)\right\rangle_{(0)}&=&\frac{\d}{{\rm{i}}\d J(1)} \F[J,\bar{J}](2)\nonumber\\
&=& \frac{ 4 {\rm{i}}m\d_{S}(1,2)  }{\square + m^{2}},\label{4b}\\
\left\langle T \F(1)\fb(2)\right\rangle_{(0)}&=&\frac{\d}{{\rm{i}}\d J(1)} \fb[J,\bar{J}](2)\nonumber\\
&=& -\frac{  {\rm{i}}D^{2}_{2}\d_{S}(1,2)  }{\square + m^{2}}\nonumber\\
&=& -\frac{  {\rm{i}}\bar{D}^{2}_{1}\d_{\bar{S}}(1,2)  }{\square + m^{2}},\label{5b}
\ea
and
\be
\left\langle T \fb(1)\fb(2)\right\rangle_{(0)}= \frac{ 4 {\rm{i}}m\d_{\bar{S}}(1,2)  }{\square + m^{2}}.\label{6b}
\end{equation}
Having defined the WZ-model at the tree level with its superfield propagators we are now able to discuss radiative corrections at the one loop level with the help of the Gell-Mann Low formula in terms of superfields.

\section{One loop self-energy corrections}
The one loop calculations for the self-energy are governed by the Gell-Mann Low formula \cite{piq}
\be
G(1,...,n)=\left\langle T\F(1)...\F(n)\right\rangle
= R\frac{  \left\langle T\F(1)...\F(n)e^{{\rm{i}}\G_{I}}\right\rangle_{(0)}  }{\left\langle T e^{{\rm{i}}\G_{I}}\right\rangle_{(0)}},\label{1c}
\end{equation}
where we use for $\G_{I}$ the ``deformed'' interaction of the form
\be
\G_{I}=  \frac{g}{48}\left[ \int dS\, \F\star\F\star\F + \int d\bar{S}\, \fb\star\fb\star\fb\right].
\end{equation}
No attempt is made to prove formula (\ref{1c}).
In our approach we define the model intuitively by (\ref{1c}) with a deformed interaction. In performing ``modified'' Wick contractions we will see that our procedure gives a meaningful result which is in agreement with a recent analysis in components \cite{ggrs}.\\
The main advantage of our superfield procedure is the fact that one gets a very compact result in form of supergraphs. In order to demonstrate the power of the superfield formulation it is sufficient to discuss one representative. For this reason we calculate the one loop graph with one chiral and one anti-chiral external leg. 
\begin{figure}
\begin{center}
\epsfig{figure=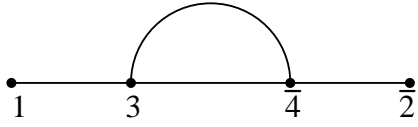,width=6cm}
\caption{Self-energy graph}
\end{center}
\end{figure}
Up to some numerical factors the corresponding contribution is given by
\be
G(1,\bar{2})\sim \left\langle T \F(1)\fb(2)\left[ \int dS_{3}\, \F^{\star 3}(3)+ \int d\bar{S}_{3}\,\fb^{\star 3}(3)\right]\left[ \int dS_{4}\, \F^{\star 3}(4)+ \int d\bar{S}_{4}\,\fb^{\star 3}(4)\right]\right\rangle_{(0)}.\label{1q}
\end{equation}
By Wick contractions in the presence of deformed interactions we calculate the graph shown in fig.1, which is one of the four contributions corresponding to\footnote{For further details in the commutative case see \cite{piq}.} (\ref{1q}). We find
\ba
G(1,\bar{2})_{3\bar{4}}&\sim&
\int\! dS_{3}\int\! \prod_{i=1}^{3}dk_{i}\,e^{{\rm{i}}\left(k_{1}+k_{2}+k_{3}\right)x_{3}}e^{-{\rm{i}}\sum_{i<j}k_{i}\times k_{j}} 
\int\! dx_{31}\int\! dx_{32}\!\int\! dx_{33}\,e^{-{\rm{i}}k_{1}x_{31}-{\rm{i}}k_{2}x_{32}-{\rm{i}}k_{3}x_{33}   }\nonumber\\&\times&
\int\! d\bar{S}_{4}
\int\! \prod_{i=1}^{3}dk'_{i}\,e^{{\rm{i}}\left(k'_{1}+k'_{2}+k'_{3}\right)x_{4}}e^{-{\rm{i}}\sum_{i<j}k'_{i}\times k'_{j}} 
\int\! dx_{41}\int\! dx_{42}\int\! dx_{43}\,e^{-{\rm{i}}k'_{1}x_{41}-{\rm{i}}k'_{2}x_{42}-{\rm{i}}k'_{3}x_{43}   }\nonumber\\&\times&
\left\langle T \F(1)\F(3)\right\rangle_{(0)}\left( \left\langle T \F(3)\fb(4)\right\rangle_{(0)} \right)^{2}\left\langle T \fb(4)\fb(2)\right\rangle_{(0)}.
\ea
The required free superfield propagators are defined in eqs.(\ref{4b})-(\ref{6b}). A straightforward but lengthy calculation leads to\footnote{We omit the ${\rm{i}}\e$-prescription. }
\be
G(1,\bar{2})_{3\bar{4}}\sim \int dp_{1}\,e^{{\rm{i}}p_{1}(x_{1}-x_{2})}\frac{1}{\left(p_{1}^{2}-m^{2}\right)^{2}}\G_{\F\fb}(p_{1}),
\end{equation}
where $\G_{\F\fb}$ is the self-energy 1PI-vertex, in the one loop approximation (up to some numerical factors) given by
\ba
\G^{(1)}_{\F\fb}(p_{1})&\sim & \int dk\,\frac{1}{2}\left[1+ \mbox{cos}(2 p_{1}\times k)\right]\frac{D^{2}_{2}(p_{1}-k)\tilde{\d}_{S}(1,2)}{(p_{1}-k)^{2}-m^{2}}\frac{D^{2}_{2}(k)\tilde{\d}_{S}(1,2)}{k^{2}-m^{2}}\nonumber\\
&=&\int dk\,\mbox{cos}^{2}\left(p_{1}\times k\right)
\frac{D^{2}_{2}(p_{1}-k)\tilde{\d}_{S}(1,2)}{(p_{1}-k)^{2}-m^{2}}\frac{D^{2}_{2}(k)\tilde{\d}_{S}(1,2)}{k^{2}-m^{2}}.\label{3c}
\ea
Using \cite{piq}
\be
D^{2}_{2}(p)\tilde{\d}_{S}(1,2)=\bar{D}^{2}_{1}(p)\tilde{\d}_{\bar{S}}(1,2)=e^{E_{12}p},
\end{equation}
one gets finally
\be
\G^{(1)}_{\F\fb}(p_{1})\sim e^{E_{12}p_{1}}\int dk\,\frac{1}{(p_{1}-k)^{2}-m^{2}}\frac{1}{k^{2}-m^{2}}\mbox{cos}^{2}\left(p_{1}\times k\right),\label{2c}
\end{equation}
where $E_{12}$ is defined by
\be
E_{12}=\q_{1}\s\qb_{1} + \q_{2}\s\qb_{2}-2\q_{1}\s\qb_{2}.
\end{equation}
The result in the form of eq.(\ref{2c}) shows in a very elegant manner that the total ``undeformed'' $\q$-dependence is encoded in the exponent $E_{12}$, whereas the remaining Feynman integral represents the ``component'' result.\\
From (\ref{3c}) it is seen that the one loop self-energy corrections are just the sum of a usual planar contribution (the '$1$' term in $[...]$ of (\ref{3c})) and a non-planar contribution (the cos(.) term in $[...]$). The planar contribution contains the expected logarithmically divergent wave function renormalization \cite{piq,ggrs}. Finally we must show that the non-planar integral
\be
I(p,\ti{p})=\int dk\,\frac{e^{{\rm{i}}k\ti{p}}}{\left(\left(p-k\right)^{2}-m^{2}+{\rm{i}}\e\right)\left(k^{2}-m^{2}+{\rm{i}}\e\right)},
\end{equation}
leads to a finite result for non-vanishing $\ti{p}^{\m}=\S^{\m\n}p_{\n}$. The calculations are given in appendix B. We find
\begin{align*}
I(p,\tilde{p}) & = -  \frac{2\mathrm{i}}{(4\p)^2}  \int_0^1 \!\!\! dx \,
K_0\Big( \sqrt{(m^2 - p^2x(1{-}x))(-\tilde{p}^2)}\Big)~.
\end{align*}


\section{Feynman rules in momentum space and one loop vertex corrections}
In order to be complete this section is devoted to discuss the one loop vertex corrections directly with the help of the Feynman rules in momentum space. With the conventions of \cite{piq} one has the following Feynman rules in momentum space \\
\unitlength 1mm
\ba
\begin{picture}(40,8)
\put(-17,-19){
\parbox{4.2cm}{
\epsfig{figure=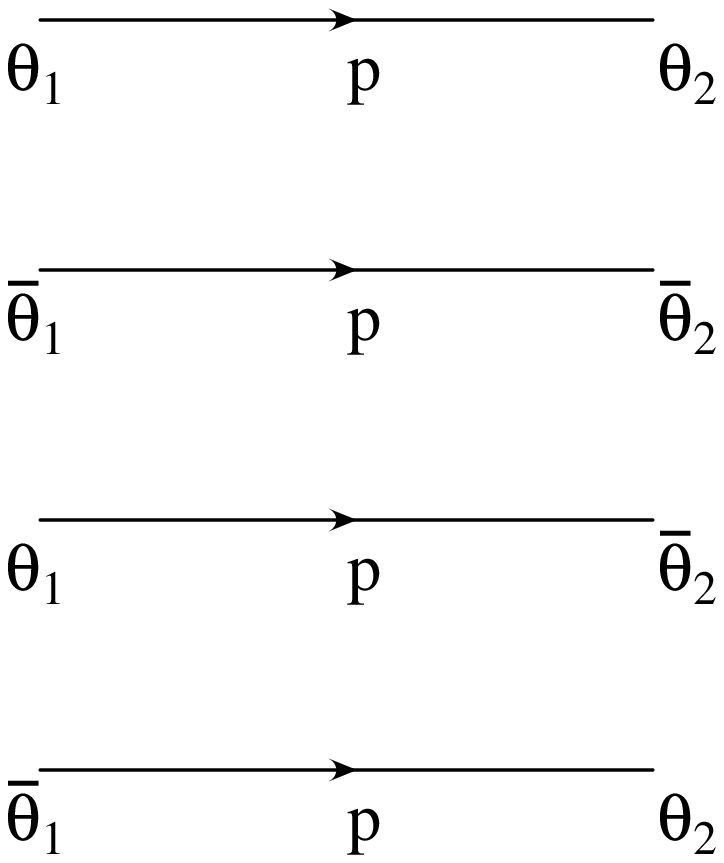,width=4.2cm}}
}
\end{picture}
\langle T\F(1)\F(2)\rangle_{(0)}^{\ti{}}&=&\frac{4{\rm{i}}m\tilde{\d}_{S}(1,2)}{p^{2}-m^{2}+{\rm{i}}\e},\label{f}\\
\langle T\fb(1)\fb(2)\rangle_{(0)}^{\ti{}}&=&\frac{4{\rm{i}}m\tilde{\d}_{\bar{S}}(1,2)}{p^{2}-m^{2}+{\rm{i}}\e},\\
\langle T\F(1)\fb(2)\rangle_{(0)}^{\ti{}}&=&\frac{{\rm{i}}D^{2}_{2}(p)\tilde{\d}_{S}(1,2)}{p^{2}-m^{2}+{\rm{i}}\e}
=\frac{{\rm{i}}\bar{D}^{2}_{1}(p)\tilde{\d}_{\bar{S}}(1,2)}{p^{2}-m^{2}+{\rm{i}}\e},\\
\langle T\fb(1)\F(2)\rangle_{(0)}^{\ti{}}&=&\frac{{\rm{i}}\bar{D}^{2}_{2}(p)\tilde{\d}_{\bar{S}}(1,2)}{p^{2}-m^{2}+{\rm{i}}\e}
=\frac{{\rm{i}}D^{2}_{1}(p)\tilde{\d}_{S}(1,2)}{p^{2}-m^{2}+{\rm{i}}\e},\label{f2}
\ea
\ba
\begin{picture}(30,10)
\put(-18,-10){
\parbox{4.2cm}{
\epsfig{figure=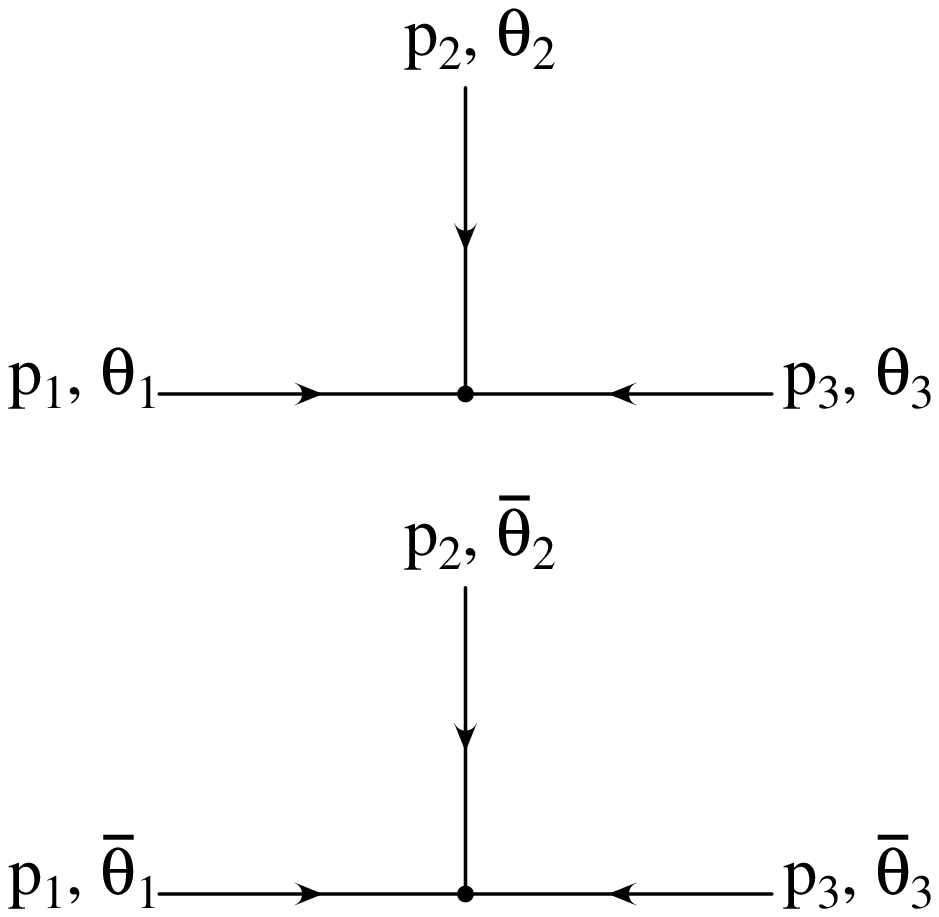,width=4.5cm}}}
\end{picture}
&&\frac{1}{8}g(2\p)^{4}\d\left(p_{1}+p_{2}+p_{3}\right)\d\left(\q_{12}\right)\d\left(\q_{13}\right)\mbox{cos}\left(p_{2}\times p_{3}\right),\\
&&\nonumber\\&&\nonumber\\&&\nonumber\\
&&\frac{1}{8}g(2\p)^{4}\d\left(p_{1}+p_{2}+p_{3}\right)\d\left(\qb_{12}\right)\d\left(\qb_{13}\right)\mbox{cos}\left(p_{2}\times p_{3}\right).\label{f1}
\ea
$ $\\
With the Feynman rules (\ref{f})-(\ref{f1}) one easily confirms the result of eq.(\ref{3c}). Additionally, one can state the non-renormalization theorem for $\G_{\F\F}$ and $\G_{\F\F\F}$ with only chiral (or anti-chiral) internal lines. A possible self-energy correction of this type is
\be
 \G_{\F\F}\sim g^2\int dk\,\frac{4{\rm{i}}m\tilde{\d}_{S}(1,2)}{(p+k)^{2}-m^{2}+{\rm{i}}\e}\frac{4{\rm{i}}m\tilde{\d}_{S}(2,1)}{k^{2}-m^{2}+{\rm{i}}\e}\mbox{cos}^2\left(p\times k\right).
\end{equation}
Due to $\tilde{\d}_{S}(1,2)\tilde{\d}_{S}(2,1)\sim \frac{1}{16}\left(\q_{12}^{2}\right)^{2}=0$ one has: $\G_{\F\F}=0$. In a similar manner one can show that also the one loop vertex correction shown in fig.2a vanishes
\begin{figure}[t]
\begin{center}
\epsfig{figure=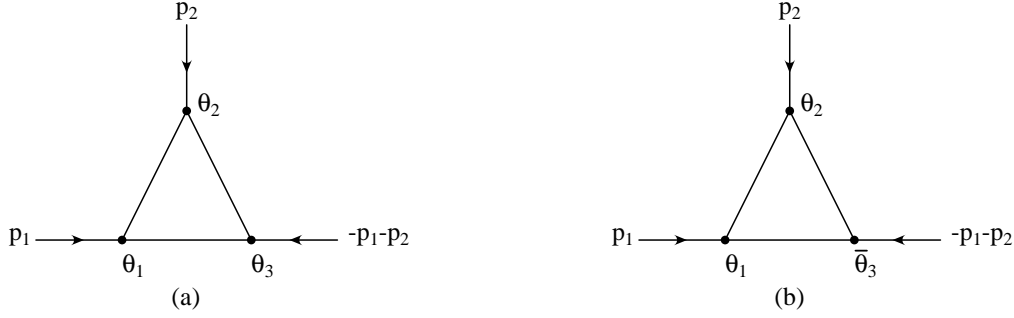,width=14cm}
\caption{One loop 3-point graphs}
\end{center}
\end{figure}
\ba
\G_{\F\F\F}&\sim&g^3 \int dk\, (4{\rm{i}}m)^{3}\frac{\tilde{\d}_{S}(1,2)}{(p_{1}+k)^{2}-m^{2}+{\rm{i}}\e}
\frac{\tilde{\d}_{S}(2,3)}{(p_{1}+p_{2}+k)^{2}-m^{2}+{\rm{i}}\e}
\frac{\tilde{\d}_{S}(3,1)}{k^{2}-m^{2}+{\rm{i}}\e}\nonumber\\
 && \times \mbox{cos}\left(p_{1}\times k\right)
\mbox{cos}\left(p_{2}\times p_{1}+p_{2}\times k\right)
\mbox{cos}\left(p_{1}\times k+p_{2}\times k\right)\label{label2}.
\ea
Using the appendix A we find that the product of the three $\ti{\d}$-functions vanishes.
However, there could exist a non-vanishing one loop correction if one allows mixed propagators, see fig.2b. Applying the above Feynman rules one gets
\ba
\G_{\F\F\fb}&\sim&  g^{3}\int dk\,
\frac{4{\rm{i}}m\tilde{\d}_{S}(1,2)}{(p_{1}+k)^{2}-m^{2}+{\rm{i}}\e}
\frac{{\rm{i}}D^{2}_{3}(p_{1}+p_{2}+k)\tilde{\d}_{S}(2,3)}{(p_{1}+p_{2}+k)^{2}-m^{2}+{\rm{i}}\e}
\frac{{\rm{i}}D^{2}_{3}(k)\tilde{\d}_{S}(3,1)}{k^{2}-m^{2}+{\rm{i}}\e}\nonumber\\&&
\times\frac{1}{4}\left[\mbox{cos}\left(p_{1}\times p_{2}\right)+\mbox{cos}\left(p_{2}\times p_{1}+2(p_{1}+p_{2})\times k\right)
+\mbox{cos}\left(p_{2}\times p_{1}-2p_{1}\times k\right)\right.\nonumber\\&&\left.
+\mbox{cos}\left(p_{2}\times p_{1}+2p_{2}\times k\right)\right]\label{label},
\ea
where we have separated the vertex correction in a planar (the $\sim\mbox{cos}\left(p_{1}\times p_{2}\right)$ term, which does not depend on the internal momentum) and a non-planar contribution. Using (\ref{a})
we conclude that the integral (\ref{label}) is finite, since it goes asymptotically like $\sim\frac{1}{k^6}$. 
 This is required in order to secure stability of the classical action.

\section{Renormalization at all orders and conclusion}

Using the same integration techniques as in appendix B one can prove
that the superficial integration encoded in any non-planar 1PI Feynman
(sub)graph $\gamma$ produces a Bessel function $K_.(.)$ depending on the
external momenta of $\gamma$. The Bessel function $K_.(.)$ tends
exponentially to zero if the external momenta of $\gamma$ become
large. This suffices to render the integration of a graph $\gamma'$
containing $\gamma$ as a subgraph UV finite. The only problem could be
an IR singularity of the Bessel function. However, since in the
commutative Wess-Zumino model there are only logarithmic divergences
(in the 2-point function), and because the only difference on the
noncommutative space are phase factors, there can only be a
logarithmic IR singularity coming from the Bessel function.
Nested logarithmic singularities are IR-integrable, as it was explicitly
demonstrated in \cite{gkw}. In conclusion, a graph in the WZ model
which contains non-planar sectors leads always to a convergent
integral. In particular, a non-planar graph in the standard sense of
the commutative world is always convergent. Divergences come only
from completely planar graphs, and they are subtracted e.g.\ by the
BPHZ procedure as in \cite{piq}. We therefore conclude that the Wess-Zumino
model on noncommutative Minkowski space is renormalizable to all loop orders, 
a result which was already obtained in \cite{ggrs} and conjectured in \cite{chr}. 
Note that the $\beta$ functions of the
noncommutative and the commutative theory differ because the
standard non-planar graphs become finite on the noncommutative space.

In this paper we have demonstrated the strength of the superfield formalism.
Especially we would like to emphasize that the superfield formalism enables us
to apply eq.(36), which lowers the degree of divergence (both IR and UV) 
by two. Furthermore is the number of graphs to be computed considerable lower
than in the work of \cite{ggrs}. We believe that this formalism will prove useful for further investigations,
in particular for super Yang-Mills theories on noncommutative $\mathbb{R}^4$.

\allowdisplaybreaks[2]
\begin{appendix}
\section{Conventions and useful formulae in superspace}
Let us briefly summarize some of the conventions and rules concerning supersymmetry and superspace (most of the rules are taken from \cite{piq}).
\subsection*{Metrics, index transport and scalar products}
The metric tensor of Minkowski space is given by $g_{\mu\nu}= \mathrm{diag}(1, -1, -1, -1)$ and we use the following spinor metric:
\begin{gather}
  \epsilon^{\alpha\beta} = \mathrm{i}\sigma^2 = 
                                    \epsilon^{\dot{\alpha}\dot{\beta}}, \\
  \epsilon_{\alpha\beta} = -\mathrm{i}\sigma^2 = 
                                    \epsilon_{\dot{\alpha}\dot{\beta}}, \\
  \epsilon_{\alpha\beta} \epsilon^{\beta\gamma} = \delta^{\gamma}_{\alpha},
\end{gather}
\begin{equation}
  \theta \eta = \theta^{\alpha} \eta_{\alpha}, \quad \bar{\theta} \bar{\eta} = 
    \bar{\theta}_{\dot{\alpha}} \bar{\eta}^{\dot{\alpha}},
\end{equation}
\begin{equation}
  \theta_{\alpha} = \epsilon_{\alpha\beta} \theta^{\beta}, \quad 
    \bar{\theta}^{\dot{\alpha}} = \epsilon^{\dot{\alpha}\dot{\beta}} \bar{
      \theta}_{\dot{\beta}}.
\end{equation}
\subsection*{Pauli matrices}
\begin{gather}
  \sigma^0 = \left( \begin{array}{cc}
                      1 & 0 \\
                      0 & 1
                    \end{array} \right), \quad
  \sigma^1 = \left( \begin{array}{cc}
                      0 & 1 \\
                      1 & 0
                    \end{array} \right), \quad
  \sigma^2 = \left( \begin{array}{cc}
                      0 & -\mathrm{i} \\
                      \mathrm{i} & 0
                    \end{array} \right), \quad
  \sigma^3 = \left( \begin{array}{cc}
                      1 & 0 \\
                      0 & -1
                    \end{array} \right),\\
  \vec{\sigma} = (\sigma^1, \sigma^2, \sigma^3), \notag \\
  \sigma^{\mu} = ({1\negthickspace\bot}_2,\vec{\sigma}) = (\sigma^{\mu})_
    {\alpha \dot{\beta}}, \notag \\
  \bar{\sigma}^{\mu} = ({1\negthickspace\bot}_2,-\vec{\sigma}) = (\bar
    {\sigma}^{\mu})_{\dot{\alpha}\beta}.
\end{gather}
%
\subsection*{Covariant derivatives}
\begin{gather}
  D_{\alpha} = \partial_{\alpha} - \mathrm{i} (\sigma^{\mu})_{\alpha
   \dot{\beta}} \bar{\theta}^{\dot{\beta}} \partial_{\mu}, \quad \bar{D}_{
     \dot{\alpha}} = - \bar{\partial}_{\dot{\alpha}} + \mathrm{i} \theta^{
       \beta} (\bar{\sigma}^{\mu})_{\dot{\alpha}\beta} \partial_{\mu}, \\
  \{ D_{\alpha}, \bar{D}_{\dot{\beta}} \} = 2 \mathrm{i} (\sigma^{\mu})_{\alpha\dot{\beta}} \partial_{\mu}, \quad \{ D_{\alpha}, D_{\beta} \} = \{ \bar{D}_{\dot{\alpha}}, \bar{D}_{\dot{\beta}} \} = 0.
\end{gather}
\subsection*{Integration}
\begin{align}
  \int dV \, \Phi &= \int d^4 \! x D^2 \bar{D}^2 \, \Phi \quad \text{for any 
    superfield $\Phi$}, \\
  \int dS \, \Phi &= \int d^4 \! x D^2 \, \Phi \quad \text{for a chiral 
    superfield $\Phi$ (i.e.\ $\bar{D} \Phi = 0$)}, \\
  \int d\bar{S} \, \bar{\Phi} &= \int d^4 \! x \bar{D}^2 \, \bar{\Phi} \quad 
    \text{for an anti-chiral superfield $\bar{\Phi}$ (i.e.\ $D \bar{\Phi} = 
      0$)}.
\end{align}
\subsection*{Functional differentiation, delta-functions and representations}
\begin{align}
  \var{1}{2} &= \delta_V(1, 2) \quad \text{for any superfield}, \\
  \frac{\d\F_{1}(1)}{\d\F_{1}(2)} &= \delta_S(1, 2) \quad \text{for a chiral superfield}, \\
  \frac{\d\fb_{2}(1)}{\d\fb_{2}(2)} &=\delta_{\bar{S}}(1,2) \quad \text{for an 
    anti-chiral superfield},
\end{align}
the numbers denoting points in superspace (for instance, $(1)$ is a shorthand-notation for $((x^{\mu})_1, (\theta_{\alpha})_1, ({\bar{\theta}}^{\dot{\alpha}})_1)$). The above delta-functions are in position space given by:
\begin{alignat}{2}
  \delta_V(1, 2) &= \frac{1}{16} \theta^2_{12} \bar{\theta}^2_{12} 
    \delta^4(x_1 - x_2), & \quad \int dV_{1} \, \Phi(1) \delta_V(1, 2) &= 
      \Phi(2), \\
  \delta_S(1, 2) &= -\frac{1}{4} \theta^2_{12} \delta^4(x_1 - x_2), & \quad 
    \int dS_{1} \, \Phi(1) \delta_S(1, 2) &= \Phi(2) \label{chiral}, \\
  \delta_{\bar{S}}(1, 2) &= -\frac{1}{4} {\bar{\theta}}^2_{12} 
    \delta^4(x_1 - x_2), & \quad \int d\bar{S}_{1} \, \bar{\Phi}(1)
      \delta_{\bar{S}}(1, 2) &= \bar{\Phi}(2), \label{achiral} 
\end{alignat}
Note that the first equalities of eq.(\ref{chiral}) and eq.(\ref{achiral}) are only valid in the chiral and anti-chiral representation, respectively.
The ``Fourier-transforms'' of the $\d$-functions in the real representation are
\begin{align}
  \tilde{\delta}_S(1, 2) &= -\frac{1}{4} \theta^2_{12}e^{-(\q_{1}\s\qb_{2}-\q_{2}\s\qb_{1})p}, 
\quad& D_{1}^{2}(p)\ti{\d}_{S}(1,2)&=e^{-E_{12}p},\label{a}\\
\tilde{\delta}_{\bar{S}}(1, 2)& = -\frac{1}{4} {\bar{\theta}}^2_{12}e^{-(\q_{1}\s\qb_{2}-\q_{2}\s\qb_{1})p},
\quad& \bar{D}^{2}_{1}(p)\ti{\d}_{\bar{S}}(1,2)&=e^{E_{12}p},
\end{align}
where
\be
E_{12}=\q_{1}\s\qb_{1}+\q_{2}\s\qb_{2}-2\q_{1}\s\qb_{2},
\end{equation}
\begin{equation}
  \theta^2_{12} = (\theta_1 - \theta_2)^2, \quad {\bar{\theta}}^2_{12} = 
    (\bar{\theta}_1 - \bar{\theta}_2)^2.
\end{equation}
 Different representations (real and (anti)chiral) of (anti)chiral superfields are connected by the following relations:
\begin{align}
  \Phi(x, \theta, \bar{\theta}) &= \Phi_1(x - \mathrm{i} \theta \sigma 
    \bar{\theta}, \theta) = e^{-\mathrm{i} \theta \sigma 
      \bar{\theta} \partial} \Phi_1(x, \theta), \\
  \bar{\Phi}(x, \theta, \bar{\theta}) &= \bar{\Phi}_2(x + \mathrm{i} \theta 
    \sigma \bar{\theta},\bar{\theta}) = e^{\mathrm{i} \theta \sigma 
      \bar{\theta} \partial} \bar{\Phi}_2(x, \bar{\theta}),
\end{align}
$1$ denoting the chiral and $2$ the anti-chiral representation; also, (anti)chiral fields have a simplified $\theta$-expansion:
\begin{align}
  \Phi_1(x, \theta) &= A(x) + \theta^{\alpha} \psi_{\alpha}(x) + \theta^{
    \alpha} \theta_{\alpha} F(x), \\
  \bar{\Phi}_2(x, \bar{\theta}) &= \bar{A}(x) + \bar{\theta}_{\dot{\alpha}} 
    \bar{\psi}^{\dot{\alpha}}(x) + \bar{\theta}_{\dot{\alpha}} \bar{\theta}^
      {\dot{\alpha}} \bar{F}(x),
\end{align}
and the covariant derivatives are given by:
\begin{alignat}{2}
  (D_{\alpha} \Phi)_1 &= \left(\partial_{\alpha} - 2 \mathrm{i} (\sigma^{\mu} 
     \bar{\theta})_{\alpha} \partial_{\mu} \right) \Phi_1, & \quad (\bar{D}_{
       \dot{\alpha}} \Phi)_1 &= -\bar{\partial}_{\dot{\alpha}} \Phi_1, \\
  (D_{\alpha} \Phi)_2 &= \partial_{\alpha} \Phi_2, & \quad (\bar{D}_{\dot
    {\alpha}} \Phi)_2 &= \left(-{\bar{\partial}}_{\dot{\alpha}} + 2 
      \mathrm{i} (\theta \sigma^{\mu})_{\dot{\alpha}} \partial_{\mu} \right) 
        \Phi_2.
\end{alignat}

\section{Calculation of the non-planar self-energy graph integral}

We are going to compute 
\[
I(p,\tilde{p}) = \lim_{\epsilon \to 0} \int \!\! \frac{d^4k}{(2\p)^4} \; 
\frac{e^{\mathrm{i}(\tilde{p}_0 k_0 -
    \vec{\tilde{p}} \cdot \vec{k})}}{ ((k_0-p_0)^2 - 
|\vec{k}-\vec{p}|^2 - m^2 +\mathrm{i}\epsilon)(k_0^2 - |\vec{k}|^2 -
m^2 +\mathrm{i}\epsilon)}~.
\]
We apply Zimmermann's $\epsilon$ trick \cite{zi}:
\begin{align*}
I(p,\tilde{p}) & = \lim_{\epsilon \to 0} 
\int \!\!\frac{d^4k}{(2\p)^4}  \; 
\frac{e^{\mathrm{i}(\tilde{p}_0 k_0 -
\vec{\tilde{p}} \cdot \vec{k})}}{ ((k_0{-}p_0)^2 
-(|\vec{k}{-}\vec{p}|^2 {+} m^2)(1{-}\mathrm{i}\epsilon))
(k_0^2 - (|\vec{k}|^2 {+} m^2)(1{-}\mathrm{i}\epsilon))}
\\
&= \lim_{\epsilon \to 0} 
\int \!\!\frac{d^4k}{(2\p)^4}  \int_0^1 \!\!\! dx \; 
\frac{e^{\mathrm{i}(\tilde{p}_0 k_0 -
    \vec{\tilde{p}} \cdot \vec{k})}}{ ((k_0^2 {-}2p_0k_0 x + p_0^2x)
  + (|\vec{k}|^2 {-}2\vec{p}{\cdot}\vec{k}x 
{+} |\vec{p}|^2 x {+} m^2)(\mathrm{i}\epsilon{-}1))^2 }
\\
&= \lim_{\epsilon \to 0} 
\int \!\!\frac{d^4k}{(2\p)^4} \int_0^1 \!\!\! dx \int_0^\infty \!\! d\alpha \; \alpha
(\epsilon'{-}\mathrm{i})^2  \\
& \hspace*{7em} \times 
e^{-\alpha(\epsilon'{-}\mathrm{i}) ((k_0^2 {-}2p_0k_0 x + p_0^2x)
+ (|\vec{k}|^2 {-}2\vec{p}{\cdot}\vec{k}x 
{+} |\vec{p}|^2 x {+} m^2)(\mathrm{i}\epsilon{-}1)) 
+ \mathrm{i}(\tilde{p}_0 k_0 - \vec{\tilde{p}} \cdot \vec{k}) }
\\
&= \lim_{\epsilon \to 0} 
\int \!\!\frac{d^4k}{(2\p)^4}  \int_0^1 \!\!\! dx \int_0^\infty \!\! d\alpha \; \alpha
(\epsilon'{-}\mathrm{i})^2 \, 
e^{\mathrm{i}(\tilde{p}_0 p_0 
- \vec{\tilde{p}} \cdot \vec{p})x }
\\
& \hspace*{4em} \times e^{-\alpha(\epsilon'-\mathrm{i})
(k_0 -p_0x -   \frac{\mathrm{i}\tilde{p}_0}{
2\alpha (\epsilon'-\mathrm{i})})^2 
-\alpha(\epsilon-\epsilon'+\mathrm{i}
+\epsilon\epsilon'\mathrm{i})|\vec{k} -\vec{p}x +
  \frac{\mathrm{i}\vec{\tilde{p}}}{2\alpha (\epsilon-\epsilon'+\mathrm{i}
+\epsilon\epsilon'\mathrm{i})}|^2 }  
\\
& \hspace*{4em} \times 
e^{-\alpha(\epsilon'-\mathrm{i})p_0^2x(1-x) 
-\alpha(\epsilon-\epsilon'+\mathrm{i}
+\epsilon\epsilon'\mathrm{i})(|\vec{p}|^2x(1-x) +m^2) 
-  \frac{\tilde{p}_0^2}{4\alpha (\epsilon'-\mathrm{i})}
-  \frac{|\vec{\tilde{p}}|^2}{4\alpha (\epsilon-\epsilon'+\mathrm{i}
+\epsilon\epsilon'\mathrm{i})} }~.
\end{align*}
For $\epsilon' < \epsilon$ we perform the Gaussian $k$ integration:
\begin{align*}
I(p,\tilde{p}) & = \lim_{\epsilon \to 0} \frac{1}{(4\p)^2} 
\int_0^1 \!\!\! dx \int_0^\infty \!\! \frac{d\alpha}{\alpha}
\Big(\frac{\epsilon'{-}\mathrm{i}}{
\epsilon-\epsilon'+\mathrm{i}+\epsilon\epsilon'\mathrm{i}}\Big)^{3/2} 
\, \mathrm{e}^{\mathrm{i}(\tilde{p}_0 p_0 
- \vec{\tilde{p}} \cdot \vec{p}) x}
\\
& \hspace*{4em} \times 
e^{-\alpha(\epsilon'-\mathrm{i})p_0^2x(1-x) 
-\alpha(\epsilon-\epsilon'+\mathrm{i}
+\epsilon\epsilon'\mathrm{i})(|\vec{p}|^2x(1-x) +m^2) 
-  \frac{\tilde{p}_0^2}{4\alpha (\epsilon'-\mathrm{i})}
-  \frac{|\vec{\tilde{p}}|^2}{4\alpha (\epsilon-\epsilon'+\mathrm{i}
+\epsilon\epsilon'\mathrm{i})} }
\end{align*}
The $\alpha$ integration yields a result
independent of $\epsilon'< \epsilon$:
\begin{align*}
I(p,\tilde{p}) & = \lim_{\epsilon \to 0}  \frac{2}{(4\p)^2}
\int_0^1 \!\!\! dx 
\Big(\frac{{-}1{-}\mathrm{i}\epsilon}{1{+}\epsilon^2}\Big)^{3/2} 
\, e^{\mathrm{i}(\tilde{p}_0 p_0 
- \vec{\tilde{p}} \cdot \vec{p}) x}
\\
& \hspace*{4em} \times 
K_0\Big( \sqrt{\big( p_0^2x(1{-}x) - (1{-}\mathrm{i}\epsilon) 
(|\vec{p}|^2x(1{-}x) +m^2)\big)
\big(\tilde{p}_0^2 - \tfrac{1{+}\mathrm{i}\epsilon}{1+\epsilon^2}
|\vec{\tilde{p}}|^2\big)} \Big)~.
\end{align*}
Now we can put $\epsilon \to0$ and switch to Minkowskian
scalar products:
\begin{align}
I(p,\tilde{p}) & = -  \frac{2\mathrm{i}}{(4\p)^2}  \int_0^1 \!\!\! dx \,
K_0\Big( \sqrt{(m^2 - p^2x(1{-}x))(-\tilde{p}^2)}\Big)~.
\end{align}
On the mass shell we have $p^2=m^2$ and $\tilde{p}^2 \leq 0$: If the
particle moves for instance in 3-direction then $\tilde{p}^2 =
- \S_{03}^2 m^2 - \sum_{i=1,2}(\S_{i3} |\vec{p}| +\S_{i0}
\sqrt{|\vec{p}|^2 +m^2})^2$.

\end{appendix}

\end{document}